# On the potential of mean force of a sterically stabilized dispersion

R. Catarino Centeno, E. Pérez and A. Gama Goicochea[*]

Instituto de Física, Universidad Autónoma de San Luis Potosí, Álvaro Obregón 64, 78000, San Luis Potosí, SLP, Mexico

## ABSTRACT

The potential of mean force (PMF) of a colloidal dispersion under various circumstances of current interest, such as varying solvent quality, polymer coating thickness, and addition of electrostatic interaction is obtained from radial distribution functions available from the literature. They are based on an implicit – solvent, molecular dynamics simulation study of a model $TiO_2$ dispersion that takes into account three major components to the interaction between colloidal particles, namely van der Waals attraction, repulsion between polymer coating layers, and a hard – core particle repulsion. Additionally, a screened form of the electrostatic interaction was included also. It is argued that optimal conditions for dispersion stability can be derived from a comparative analysis of the PMF under the different situations under study. This thermodynamics based analysis is believed to be more accessible to specialists working on the development of improved $TiO_2$ formulations than that based on the more abstract, radial distribution functions.

[*]Corresponding author. Electronic mail: agama@stanford.alumni.edu

**INTRODUCTION**

Titanium dioxide (TiO$_2$) particles dispersed in aqueous solvent constitute perhaps the most important industrial test bed for theories of colloidal stability and are also the focus of numerous experiments designed to understand the interaction between the TiO$_2$ particles and the polymeric dispersant to improve the conditions of optimal stability [1]. Among some of the most popular applications of TiO$_2$ dispersions are found in consumer goods such as architectural white, water – based paints [2], shampoo, toothpaste, and others [3]. There are also important environmental applications of titania dispersions [4].

It is known that a dispersion of TiO$_2$ particles can be kinetically stabilized by coating the particles with polymers or with polyelectrolytes [5]; other properties of the particles, such as preserving their high refractive index can be accomplished with other types of coatings, such as metallic oxides [6]. Coating the particles surface with polymers grafted onto the surfaces so as to form polymer "brushes" is an efficient mechanism of stability because there is an entropy gain if the opposite brushes overlap, which is thermodynamically unfavorable. Combining this mechanism with electrostatic repulsion results in an even better means of stability.

The basic interactions that compete in the phenomenon of colloidal stability are the short – range, van der Waals attraction and long – range electrostatic repulsion. Those are the basic ingredients of the so – called DLVO theory (named after the initials of Derjaguin, Landau, Verwey and Overbeek) [7], which has met with considerable success. However, van der Waals attraction is important only when the particles are not coated and can get in close contact with one another. When a polymer brush is grafted onto the particles surface, other interactions come into play, not only of entropic nature, but arising also from three – body repulsion between polymer chains [8]. These interactions have been used in the past as mechanisms to promote entropic (steric), electrostatic colloidal stability, or a combination of both [5]. Advances have been achieved through the application of density functionals and integral equations for cases such as varying ionic strength or solvent quality [9, 10]. A relatively modern alternative to the theoretical and experimental efforts devoted to the understanding and optimization of colloidal dispersion comes from the field of computer simulations [11]. Among their advantages is the fact that one can solve the interaction model for many particles almost exactly, which most theoretical approaches cannot accomplish. Also, one has total control over the thermodynamic and physicochemical conditions of the model dispersion, something that is not easily achieved in most experiments. From molecular dynamics simulations one can obtain correlation functions that can shed light on the kinetic or thermodynamic stability conditions of the dispersion. One of those functions is the pair distribution function, also known as the radial distribution function [11], which is commonly used to determine the relative spatial correlations between particles under a given interaction model. Although much has been learned over the years from the considerable amount of work amassed during that period, the theoretical and computational information remains relatively inaccessible to most researcher carrying out experiments to improve stability of formulations, because properties such as correlation functions are not as easily grasped as are thermodynamic concepts.

In this work our focus is on illustrating how some simple guidelines can be followed to use the functions mentioned above in the search for colloidal stability criteria, and apply them to a specific example taken from the literature. Additionally, we compare the PMF obtained from other competing theories (Alexander – de Gennes [12], and Milner – Witten – Cates

[13]) based on different assumptions so that a criterion can be established to uniquely determine the physical basis for colloidal stability.

**MODELS AND METHODS**

Our starting point is a mean – field theory, proposed by Zhulina and coworkers [8], hereafter referred to as ZBP, for the interaction between colloids covered with polymer brushes. Such interaction has two contributions: a short – range attractive term, arising from the van der Waals interaction between colloids, which is known to depend inversely proportionally to the distance between the colloids' surfaces [7]:

$$U_{vdW} = -\frac{A_H}{12}\frac{R}{h^2} \quad (1)$$

where $A_H$ is Hamaker's constant, $R$ is the colloidal particles' radius, and $h$ is the colloids' surface to surface distance. The other term is a repulsive contribution arising from the interaction between the polymer brushes on opposite colloidal surfaces as they approach each other, for relative separation distances that are smaller than the particles size [14]. The total interact is shown in equation (2):

$$U(r) = \Delta F_0^\theta \left[\frac{\beta\pi}{r} + \frac{\pi^2}{12r^2}(1-\gamma)\right] + U_{hc} \quad (2)$$

where $r$ is the relative distance between the colloids centers of mass, $\beta$ is the solvent's quality, $\Delta F_0^\theta$ is the free energy of the uncompressed polymer layer at the $\theta$ - temperature, and $\gamma$ is a constant that incorporates the polymer – polymer repulsive interactions through the dimensionless third virial coefficient ($\omega$), the polymer grafting density on the colloidal surface ($\Gamma$), and Hamaker's constant, namely:

$$\gamma = \frac{A_H}{96\pi k_B T \omega}\left(\frac{1}{\Gamma N a^2}\right)^3 \qquad (3)$$

with $k_B$ being Boltzmann's constant, $T$ the absolute temperature, $N$ the polymers' degree of polymerization, and $a$ the monomers' size [8]. The last term in equation (2), $U_{hc}$ is only a hard core potential whose purpose is avoiding that particles completely penetrate each other [14]. The dimensionless polymer grafting density is $\Gamma^*=\Gamma N a^2$. In Figure 1 we show the behavior of the interparticle potential shown in equation (2) for two cases: when the constant $\gamma$ in equation (3) is $\gamma < 1$, indicative of weak interparticle attraction, and $\gamma > 1$, which occurs for strong interparticle attraction.

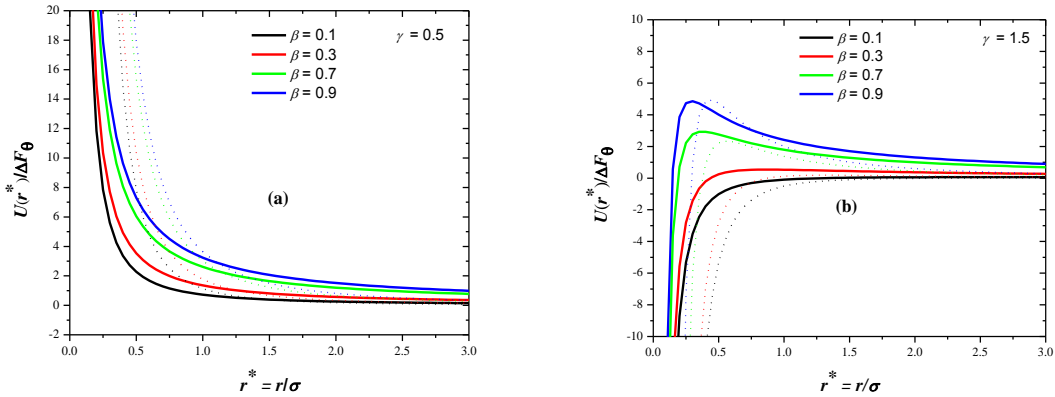

**Figure 1**. Interparticle interaction (solid lines) of colloids covered with polymer brushes, at increasing values of the solvent quality, $\beta$, from equation (2) for two values of the constant $\gamma$ in equation (3), for (a) $\gamma < 1$, and (b) $\gamma > 1$. The dotted lines represent the forces obtained from the negative derivative of the interaction function. The axes are shown in reduced units, with $\sigma$ being the diameter of the colloids and $\Delta F_\theta$ the free energy of the uncompressed polymer brush at the $\theta$ – temperature ($\beta=0$).

As shown in Figure 1 (a), for situations when $\gamma$ is less than 1, the colloidal dispersion is always stable (whenever $U(r)$ is positive), and the stability improves as the solvent quality itself is improved (increasing $\beta$). Under $\theta$ – to good – solvent conditions, the dispersion is stable because there is a dominance of three – body repulsion ($\omega$) over the van der Waals interaction, which is sufficient to make the dispersion thermodynamically stable. Pair interactions contribute to improve the stability when $\beta > 0$. When $\gamma > 1$ (see Figure 1(b)), there appears a transition from an unstable dispersion (see for example the black line in Figure 1(b)) to a stable one as the solvent quality increases (increasing $\beta$, see for example the blue line in Figure 1(b)), established by the appearance of a maximum in the interaction potential. The competition between the attractive van der Waals interaction and the repulsive three – body correlations is responsible for such maximum, which leads to a kinetically stable colloidal dispersion.

The model shown in equation (2) takes into account the three – body repulsion between monomers that make up the polymer chains (through the third virial coefficient, $\omega$), and radial distribution functions obtained through molecular dynamics computer simulations using this model [14] have demonstrated that it can lead to colloidal stability, that is repulsion between polymer – coated colloids. Other models for the effective force between polymer brushes, such as that of Alexander and de Gennes (AdG) [12] do not take into account chain – chain interaction. In particular, AdG's model assumes that the chains density profile is a step function with all the chains ends placed at the layer surface, that there is no interchain interaction, and that the polymer brushes are in a good solvent. It considers only two principal contributions to the many – body force in compressed polymer brushes: a short range repulsion, due to the osmotic pressure that arises from the increased density of monomers in the compressed region, and a medium range attraction whose origin is the elastic energy of the polymer chains. Although both models (ZBP and AdG) predict repulsion between strongly compressed polymer brushes, the physical origin of such repulsion is different. While AdG attribute it to an entropy gain when opposing chains are disordered by the compression, ZBP adjudicate it to three – body interactions between the monomers making up the chains. An alternative, self –consistent field model [13], proposed by Milner, Witten and Cates (MWC) does take into account inter chain interaction in the brush, but yields a PMF curve that differs very little from that of AdG´s. For the sake of posterior comparison, we shall consider the following expressions for the interactions between polymer brush – coated colloidal particles in good solvent with the present model:

$$\frac{W_{AdG}(h)}{k_B T} = (2h_0) A \Gamma^{3/2} \left[ \frac{4}{5} \left( \frac{2h_0}{h} \right)^{5/4} + \frac{4}{7} \left( \frac{h}{2h_0} \right)^{7/4} \right], \quad (4)$$

$$\frac{W_{MWC}(h)}{k_B T} = (2h_0) A \Gamma^{3/2} \left[ \frac{1}{2} \left( \frac{2h_0}{h} \right) + \frac{1}{2} \left( \frac{h}{2h_0} \right)^2 - \frac{1}{10} \left( \frac{h}{2h_0} \right)^5 \right]. \quad (5)$$

Equations (4) and (5) represent the PMF for the AdG and MWC models, respectively, where in both cases $h_0$ represents the thickness of the uncompressed polymer layer, $A$ is the colloids surface area, $\Gamma$ is the polymer grafting density, and $h$ is the distance separating the surfaces of colloids when the polymer layers are compressed. It must be noticed that both models are valid only for compressed polymer brushes, i. e., for $h \leq 2h_0$. Figure 2 shows a schematic diagram of the model colloidal dispersion that is the purpose of this work, for illustrating purposes only.

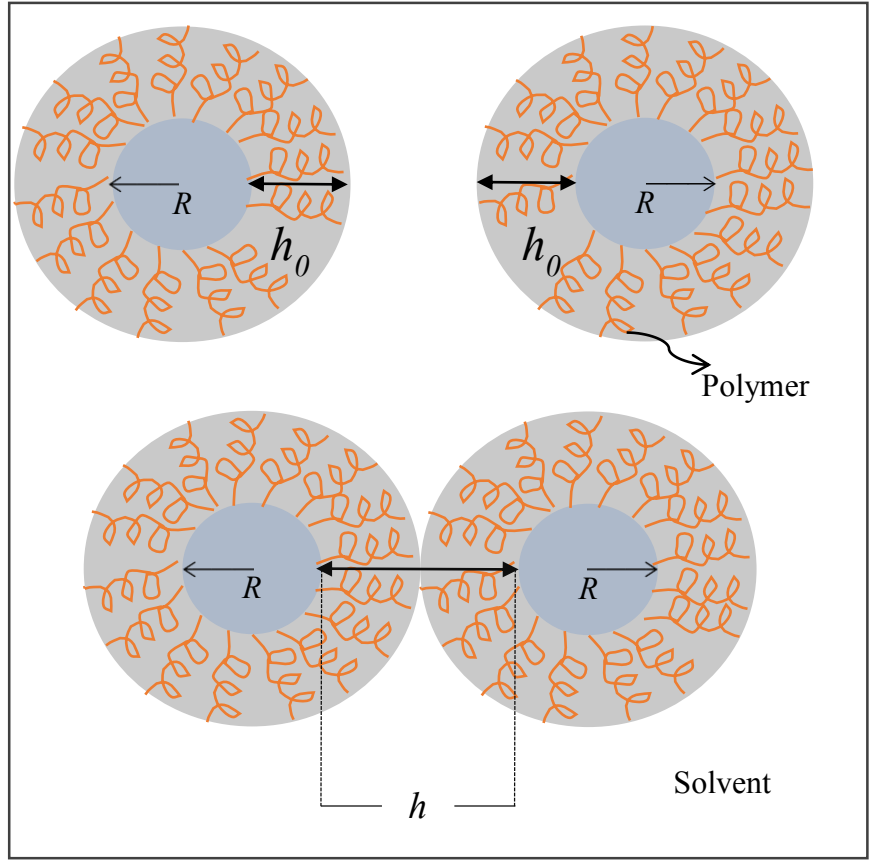

**Figure 2.** Schematic diagram of the dispersion whose stability is determined in this work assuming the colloidal particles (solid gray circles) are covered by a layer of polymer chains (in beige) grafted onto their surface, immersed in a solvent.

We focus here on the PMF ($W_{PMF}(r)$), which is an effective interaction that provides important thermodynamic information about a many – body system. It can be obtained from the colloids' radial distribution functions, $g(r)$, through the relation [15]:

$$W_{PMF}(r) = -k_B T \ln[g(r)]. \tag{6}$$

We shall use equation (6) to obtain the PMF for the ZBP model, using the radial distribution functions calculated in reference [14], and compare with the models shown in equations (4) and (5).

**RESULTS AND DISCUSSION**

First we show the PMF for ZBP's model [8] as a function of the polymer brush thickness, then for increasing quality of the solvent, and finally for a colloidal dispersion with, and without electrostatic interactions. For all cases we chose $T$=300 K; quantities expressed in reduced units are indicated with asterisks.

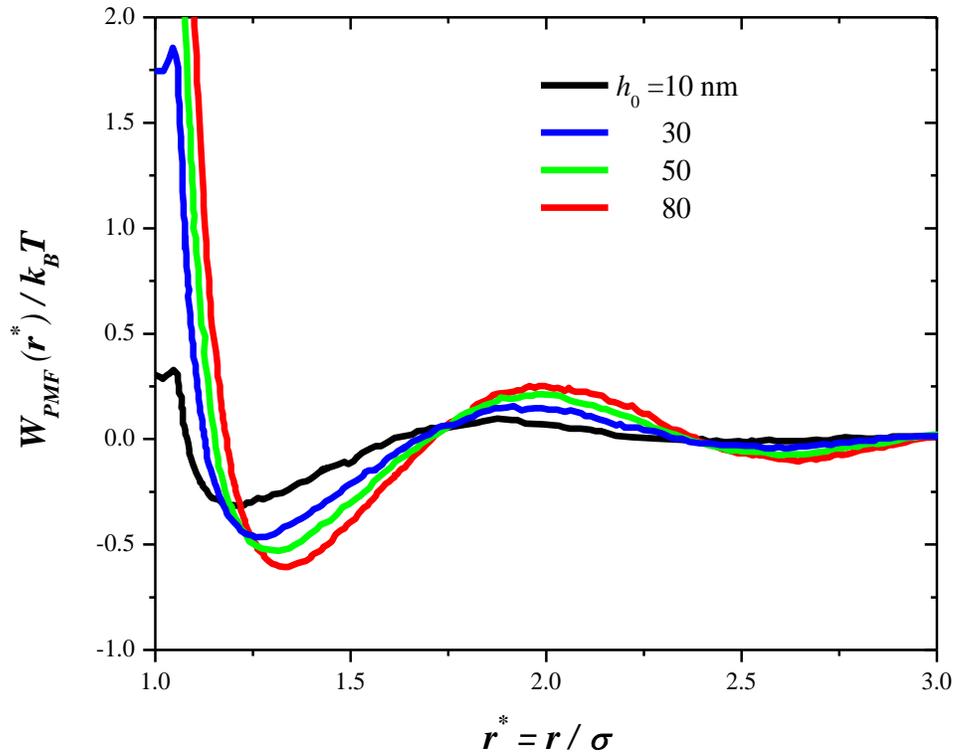

**Figure 3**. PMF for a system of colloidal particles covered with polymer brushes, of increasing thickness ($h_0$, expressed in nm), as a function of the dimensionless relative distance between the centers of mass of the particles, $r^*$, for a reduced grafting density $\Gamma^*=0.25$. The value of the Hamaker constant was chosen as $A_H=6\times10^{-20}$ J, $\sigma=200$ nm, and $\omega=1$. Increasing $h_0$ leads to better stability.

Figure 3 shows the PMF results for a TiO$_2$ colloidal dispersion in water whose Hamaker constant is $A_H=6\times10^{-20}$ J [16], with average particle size $\sigma=200$ nm [17]. For the thinnest polymer coatings (black and blue lines, with $h_0=10$ nm and 30 nm, respectively) the PMF barrier is not too high, less than $2k_BT$, and could be overcome in particle – particle collisions due to thermal fluctuations, leading to flocculation of some particles whose fraction is reduced as $h_0$ is increased, as expected [5], see the green and red lines in Figure 3. A first minimum in the PMF appears at relative distances below $r^*\sim1.5$, which is however relatively shallow and can easily be overcome by Brownian motion, followed by a second one at $r^*\sim2.7$ which is even shallower. The oscillations shown by the PMF arise from the corpuscular nature of the dispersion [7], with a period given approximately by the particle size ($\sigma$), which is considered to be monodispersed throughout this work. In actual water – based paints TiO$_2$ is known to have a distribution of sizes [18], and in such case the oscillations shown in Figure 3 are expected to be washed out, yielding a monotonically decreasing PMF.

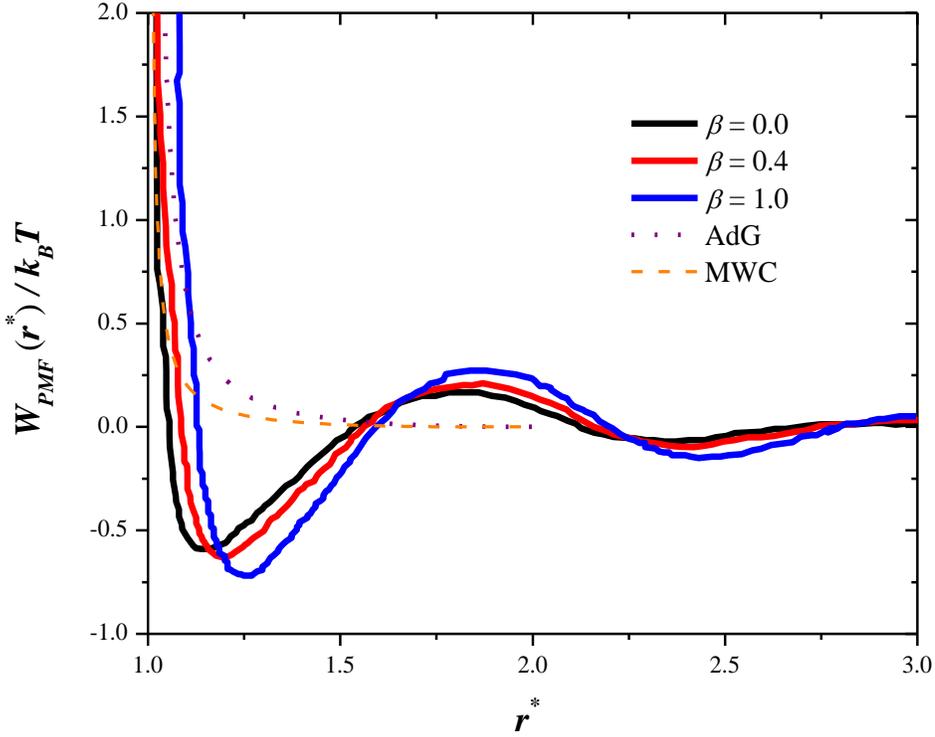

**Figure 4**. PMF for a system of colloidal particles covered with polymer brushes, of increasing solvent quality ($\beta$, solid lines), as a function of the dimensionless relative distance between the centers of mass of the particles, $r^*$, for $\gamma =0.5$ (see equation (3) and Figure 1(a)). The AdG (dotted purple line, equation (4)) and MWC (dashed orange line, equation (5)) models are included also, for comparison. The latter models are defined for compressed brushes only, becoming zero when the brushes do not overlap, which in the scale of the figure occurs at $r^* \geq 2$. See text and the Appendix for details.

Increasing the parameter $\beta$ is equivalent to improving the solvent quality, see Figure 1, with $\beta =0$ representing a theta − solvent [8]. In Figure 4 we present the PMF for a TiO$_2$ dispersion with $\gamma$ fixed at 0.5, which means stability is always obtained through ternary repulsion overcoming the van der Waals attraction (Figure 1(a)). As Figure 4 shows, inclusion of binary interactions through the solvent quality parameter, $\beta$, improves the stability even more, leading to increasingly large potential barriers, well above the thermal energy, $k_BT$. The depth of the short range ($1 < r^* < 1.5$) and larger range ($2 < r^* < 2.5$) wells increase also with $\beta$, but their depth is less than the thermal energy and therefore do not yield permanent particle flocculation. We have included in Figure 4 the PMF curves obtained from the AdG (dotted purple line) [12] and MWC (dashed orange line) [13] models, using equations (4) and (5), respectively, for comparison. To do so one has to properly normalize these equations, which is easily done if length is reduced with $2h_0$, yielding a reduced colloid area $A^*=A/(2h_0)^2$ and reduced grafting polymer density $\Gamma^*=\Gamma(2h_0)^2$. In the Appendix we show in detail how the expressions for the AdG and MWC models are reduced. Those models are defined only for compressed polymer brushes, therefore they become identically zero when the brushes do not overlap. Regarding the oscillations shown by the PMF (and the lack of them for the AdG and MWC models) in Figure 4, the same analysis as that of Figure 3 applies here. At the strongest compression of the polymer layers (for values of $r^*$ close to 1) and for good solvent conditions ($\beta=1$), the PMF (blue line in Figure 4) obtained from the ZBP model [8, 14] is more repulsive than those corresponding to the AdG and MWC models, indicating that ternary interactions should not be neglected at large compression because they can be the leading repulsive mechanism. Recent explicit – solvent computer simulations of planar surfaces coated with relatively short polymer brushes [19] have confirmed that both AdG and MWC models reproduce fairly well the PMF at intermediate compression of the brush, but they are not as good at very strong compression, as Figure 4 shows. Although ZBP was designed for colloidal particles coated with polymer brushes, it can also be applied to colloids coated with layers of adsorbed polymers (sometimes called "surface modifying" polymers [20]), because ternary interactions play the same role in both situations.

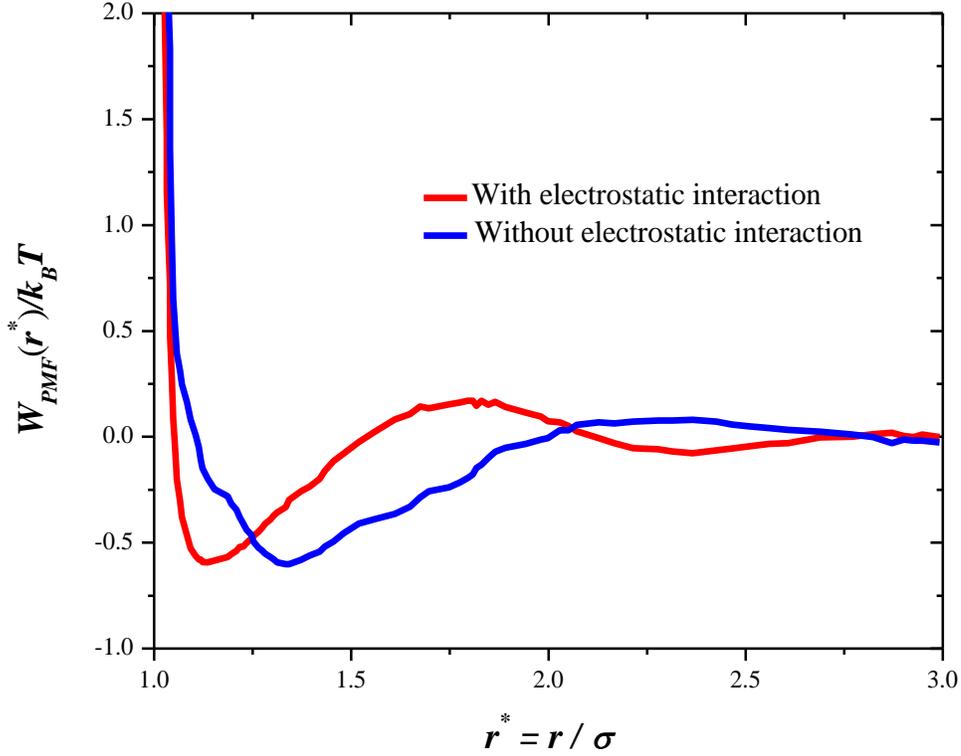

**Figure 5**. PMF for colloidal dispersions with (red line) and without (blue line) electrostatic interactions. The non – electrostatic case corresponds to a theta solvent ($\beta=0$) with $\gamma=1.05$ (see equation (2)).

Lastly, we consider briefly the influence of weak electrostatic interactions in the stability of the ZBP model. To do so in a way that is consistent with the mean – field nature of the electrically neutral model, one adds a screened electrostatic contribution to the total particle – particle interaction potential, see equation (2), of the so – called Yukawa type given by the following expression:

$$U(r) = \frac{(z_{effec}e)^2 exp(-\kappa_D r)}{(4\pi\varepsilon_0\varepsilon_r)r}. \tag{7}$$

In equation (7) $Z_{effec}$ is the effective charge on the colloidal particles surface, $e$ is the electron charge, $\kappa_D$ is the inverse of the Debye length, and $\varepsilon_0\varepsilon_r$ is the medium permittivity, and values appropriate for a TiO$_2$ dispersion in aqueous solvent have been used, see reference [14] for full details. Figure 5 shows the PMF (red line) obtained from the radial distribution function after molecular dynamics simulations are run for a system of particles where equation (7) is added to the ZBP interaction potential, equation (2). For comparison, the PMF of an electrically neutral dispersion in a theta solvent is included in the same figure. Two features are of notice in Figure 5; on the one hand, the range of the first shallow attractive well (where the PMF is negative) is reduced by about half when the electrostatic interaction is included, namely the range for the neutral dispersion (blue line in Figure 5) comprises values of $r^*$ that go from about $r^*$ ~1.2 up to $r^*$ ~2.0, while for the electrostatic case such range goes from $r^*$ ~1.1 up to $r^*$ ~1.5. On the other hand, a potential barrier appears at $r^*$ ~1.8 when electrostatics is included although it is relatively weak, amounting to about $0.17k_BT$, nevertheless this may be enough for kinetic stabilization of the dispersion. These features and the shape of the red curve in Figure 5 are reminiscent of the DLVO potential [7]. Although the PMF for the charged ZBP model shown in Figure 7 (red curve) is similar in shape to that typical of the DLVO model, they arise from competition of forces of different origin. While DLVO is the result of attractive van der Waals interactions coexisting with repulsive double – layer electrostatic interactions, the charged version of the ZBP model produces a barrier in the PMF whose origin is the competition between attractive van der Waals interactions, ternary repulsion between polymer brushes, and repulsive electrostatic interactions between the colloidal surfaces. For high grafting density and large polymerization degree the amplitude of the oscillations that the curves in Figure 5 show is expected to be reduced. However, the PMF does have a strong dependence on the electrostatic properties of the dispersion, such as the polyelectrolytes pH, or ionization degree, becoming more repulsive as the ionization degree increases [21]. Although it is not the purpose of this work to carry out a systematic study of the influence of such factors on the PMF, it is important to emphasize that the inclusion of electrostatics in the model leads to trends in the PMF that can be interpreted as those of increasing the polymer layer thickness, or the quality of the solvent.

**CONCLUSIONS**

Optimizing the stability of colloidal dispersions for current applications requires the use of sophisticated methods that go beyond the traditional trial and error experimental tests. One of such methods is the application of molecular dynamics simulations, which can be run on modern computers to yield important physicochemical information directly comparable with experiments, in a relatively short time. Although such methods have met with success when applied to paints and coatings [22], they still remain relatively inaccessible to these communities due to in part to a lack of familiarity with properties routinely obtained from molecular dynamics simulations, such as the radial distribution functions. Here we show that a clearer understanding of the conditions for stability of a colloidal dispersion can easily be obtained from analysis of the PMF, which can be directly obtained from radial distribution functions and reduces to thermodynamics analysis. The colloidal stability of the sterically stabilized dispersions is clearly dependent on the solvent quality, polymer coating thickness, and strength of the electrostatic interaction. The latter is important but not necessary for the colloidal stability of these systems because polymer brushes are able to produce a ternary repulsion among these polymers. The van der Waals attraction between particles is then balanced by this repulsion producing a PMF that is very similar in shape to the DLVO potential but without the presence of the electrostatic repulsion. Analysis of the PMF showed that increasing the thickness of the polymer layer that coats the colloidal particles, improving the quality of the solvent (which can be done raising the temperature) or adding charges to the system have the same effect that is, improving the kinetic stability of the colloidal dispersion.

**ACKNOWLEDGEMENTS**

AGG would like to thank Universidad Autónoma de San Luis Potosí for the hospitality and necessary support for this project. The authors acknowledge M. A. Balderas Altamirano for discussions.

**APPENDIX**

Here we show how equations (4) and (5) can be expressed in reduced units (indicated by an asterisk) so that they can be drawn on the same scale as the other PMF in Figure 3. Let us start by changing variables so that the spatial coordinate is not the compressed polymer layer thickness ($h$) but rather the relative distance between the colloidal particles centers of mass ($r$), as follows: $r = h + \sigma$, where $\sigma$ is the particles diameter (see Figure 2). Reducing all lengths with $2h_0$, we get for the AdG model

$$W^*_{AdG}(r^*) = \frac{W_{AdG}(r^*)}{k_B T} = A^* \Gamma^{*3/2} \left[ \frac{4}{5}(r^* - \delta^*)^{-5/4} + \frac{4}{7}(r^* - \delta^*)^{7/4} - \frac{48}{35} \right], \tag{A1}$$

and for MWC model

$$W^*_{MWC}(r^*) = \frac{W_{MWC}(r^*)}{k_B T} = A^* \Gamma^{*3/2} \left[ \frac{1}{2}(r^* - \delta^*)^{-1} + \frac{1}{2}(r^* - \delta^*)^2 - \frac{1}{10}(r^* - \delta^*)^5 - \frac{9}{10} \right],$$

(A2)

where $A^* = \pi \delta^{*2} A$; $\Gamma^* = N_p/A^*$, with $N_p$ equal to the number of polymer chains grafted onto the colloidal surface, and $\delta^* = \sigma/2h_0$. The constants subtracted (48/35 in equation (A1), and 9/10 in equation (A2)) are chosen so that the PMF be equal to zero when the opposing polymer brushes separate enough that they do not overlap ($r^* = 2\delta^*$), since both models (AdG [12] and MWC [13]) are defined only for polymer brush compression. To compare both models with our predictions for the PMF on the same scale we chose the value of $A^* \Gamma^{*3/2} = 0.05$ and $\delta^* = 1$, for both cases (equations (A1) and (A2)).